\documentclass[3p,times,twocolumn]{elsarticle}

\usepackage{ecrc}
\usepackage{amsmath}
\usepackage{amssymb}
\usepackage[figuresright]{rotating}
\usepackage{multirow}

\volume{00}

\firstpage{1}

\journalname{Nuclear and Particle Physics Proceedings}

\runauth{}

\jid{nppp}

\jnltitlelogo{Nuclear and Particle Physics Proceedings}

\begin{document}

\begin{frontmatter}

\dochead{}

\title{Meson cloud effects on kaon quark distribution functions and the SU(3) flavor symmetry$^*$}

\cortext[cor0]{Talk given at 21st International Conference in Quantum Chromodynamics (QCD 18), July 2 - 6, 2018, Montpellier - FR}
\author[label1,label2]{Akira Watanabe\fnref{fn1}}
\fntext[fn1]{Speaker, Corresponding author.}
\ead{akira@ihep.ac.cn}
\address[label1]{Institute of High Energy Physics and Theoretical Physics Center for Science Facilities, Chinese Academy of Sciences, Beijing 100049, People's Republic of China}
\address[label2]{University of Chinese Academy of Sciences, Beijing 100049, People's Republic of China}
\author[label3]{Takahiro Sawada}
\ead{tsawada@sci.osaka-cu.ac.jp}
\address[label3]{Department of Physics, Osaka City University, Osaka 558-8585, Japan}
\author[label4]{Chung Wen Kao}
\ead{cwkao@cycu.edu.tw}
\address[label4]{Department of Physics and Center for High Energy Physics, Chung-Yuan Christian University, Chung-Li 32023, Taiwan}

\begin{abstract}
We present our analysis on the valence quark distribution functions of the $K^{+}$ meson,
$v^{K (u)} (x, Q^2)$
and
$v^{K (\bar{s})} (x, Q^2)$,
taking into account the meson cloud effects in the framework of the chiral constituent quark model.
Assuming appropriate bare quark distribution functions at the initial scale, evaluating the Goldstone boson dressing corrections with the model, and performing the QCD evolution, one can obtain the realistic dressed distributions at a certain scale.
The phenomenologically satisfactory valence quark distribution of the pion was obtained in the previous study and there are available experimental data of the valence $u$ quark distribution ratio
$v^{K(u)} (x, Q^2) / v^{\pi (u)} (x, Q^2)$,
which enable us to obtain the kaon quark distribution functions.
Besides presenting the resulting dressed distributions, we also show how the meson cloud effects affect the bare distributions in detail.
A striking result is that the observed size of the SU(3) flavor symmetry breaking is considerably smaller than those seen in the preceding works based on other approaches.
\end{abstract}


\end{frontmatter}

%%%%%%%%%%%%%%%%%%%%%%%%%%%%%%%%%%%%%%%%
\section{Introduction}
\label{sec-1}
%%%%%%%%%%%%%%%%%%%%%%%%%%%%%%%%%%%%%%%%
Understanding the quark-gluon structure of hadrons is one of the most important research subjects in hadron physics, and a tremendous number of experimental and theoretical studies has been done over several decades.
However, it is also true that most of the preceding studies were devoted to the nucleon structure, and our knowledge about the meson structure, particularly for the Goldstone bosons, is still quite limited.
Thanks to the recent developments in the experimental technique, currently the pion-induced Drell-Yan experiment has been performed by the COMPASS collaboration at CERN.
Moreover, the kaon-induced Drell-Yan experiment might also be realized in the future, which would provide us with valuable opportunities to experimentally investigate the kaon structure.
Due to such a situation, improvements of the theoretical understandings are also expected.

The partonic structure of a hadron is encoded into the parton distribution functions (PDFs).
Although the fundamental theory of the strong interaction, quantum chromodynamics (QCD), is established, one cannot directly calculate PDFs because of the nonperturbative nature of a hadron.
Hence, effective approaches are necessary to study them.
In this work, we investigate the valence quark distribution functions of the kaon, and one of the most suitable models to achieve this is the chiral constituent quark model~\cite{Melnitchouk:1994en,Kulagin:1995ia}.
In the model, the valence quarks are described as the constituent quarks which are surrounded by the meson clouds.
The dressing corrections by the clouds can be evaluated within the model, assuming the bare distribution functions for the constituent quarks at the initial scale $Q_0^2$.
The dressed distributions obtained at the low scale are evolved by the Dokshitzer-Gribov-Lipatov-Altarelli-Parisi (DGLAP) equation to a higher scale $Q^2$, at which the experimental data for various high energy scattering processes are usually taken.
In this framework, there is no gluon and no sea quark at $Q_0^2$, but they emerge via this QCD evolution process.
So far, this model was applied to the analysis of the nucleon~\cite{Suzuki:1997wv} and the pion~\cite{Watanabe:2016lto} structures.

In this report, based on Ref.~\cite{Watanabe:2017pvl} we present our recent analysis on the kaon quark distribution functions in the framework of the chiral constituent quark model.
We assume the functional forms of the bare quark distributions, which include some adjustable parameters, at the initial scale, and evaluate the dressing corrections.
The dressed quark distributions are evolved to the higher scale, and then we compare the resulting distributions with the experimental data of the valence $u$ quark distribution ratio of the pion and the kaon~\cite{Badier:1980jq} to determine the parameters.
Since the bare distribution of the pion is required for this procedure, we utilize the results of the previous study~\cite{Watanabe:2016lto}.
After we explain the formalism and present the analytical expressions for the dressed quark distributions in the next section, we will show the numerical results for the resulting valence quark distribution functions of the kaon, from which the size of the SU(3) flavor symmetry breaking can be seen.
We will also show how the meson cloud effects affect the bare quark distributions.

%%%%%%%%%%%%%%%%%%%%%%%%%%%%%%%%%%%%%%%%
\section{Dressing corrections to constituent quarks}
\label{sec-2}
%%%%%%%%%%%%%%%%%%%%%%%%%%%%%%%%%%%%%%%%
Here we briefly explain the formalism to evaluate the dressing corrections, and then present the resulting analytical expressions for the dressed quark distributions.
The fields of the constituent quarks and the Goldstone bosons are defined as
\begin{equation}
\psi =
\begin{pmatrix}
u \\
d \\
s \\
\end{pmatrix}
, \ \
\Pi = \frac{1}{\sqrt{2}}
\begin{pmatrix}
\frac{\pi^{0}}{\sqrt{2}}+\frac{\eta}{\sqrt{6}}
&
\pi^{+}
&
K^{+} \\
\pi^{-}
&
-\frac{\pi^{0}}{\sqrt{2}} + \frac{\eta}{\sqrt{6}}
&
K^{0} \\
K^{-}
&
\bar{K}^{0}
&
-\frac{2 \eta}{\sqrt{6}} \\
\end{pmatrix}
,
\end{equation}
respectively.
Their interactions are described by the effective Lagrangian:
\begin{equation}
{\mathcal{L}}_{int} = - \frac{g_A}{f} \bar{\psi} \gamma^{\mu}\gamma_{5} (\partial_{\mu} \Pi) \psi ,
\end{equation}
where $g_A$ and $f$ represent the quark axial-vector coupling constant and the pseudoscalar decay constant, respectively.

\begin{figure}[tb!]
\centering
\includegraphics[width=0.45\textwidth]{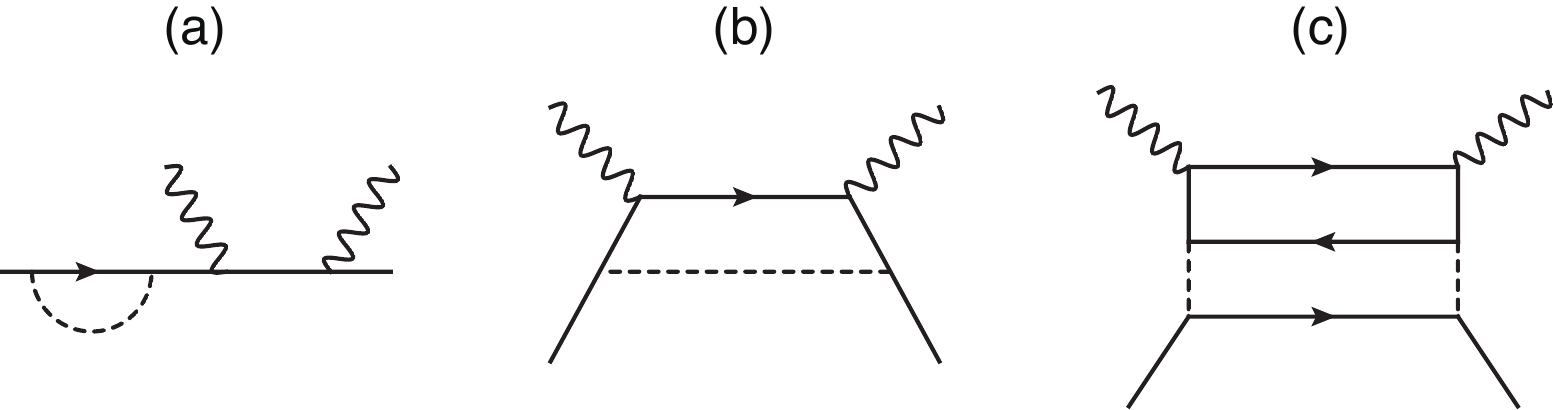}
\caption{
The diagrams representing the dressing corrections to a constituent quark.
The thick, wavy, and dashed lines depict the constituent quarks, the virtual photons, and the Goldstone bosons, respectively.
}
\label{fig:dressing_corrections}
\end{figure}

In this study, we take into account the dressing corrections described by the three diagrams in Fig.~\ref{fig:dressing_corrections}.
The contributions from the renormalization constants are depicted by Fig.~\ref{fig:dressing_corrections}(a).
The virtual photon directly couples to the constituent quark in Fig.~\ref{fig:dressing_corrections}(b), but the photon probes the Goldstone boson structure in Fig.~\ref{fig:dressing_corrections}(c).
To evaluate the contributions from those diagrams, we firstly introduce the splitting function:
\begin{equation}
{P_{j \alpha / i}}\left( x \right)  =  \frac{1}{{8{\pi ^2}}}{\left( {\frac{{{g_A} \bar m}}{f}} \right)^2}\int {dk_T^2} \frac{{{{\left( {{m_j} - {m_i} x} \right)}^2} + k_T^2}}{{{x^2}\left( {1 - x} \right){{\left( {m_i^2 - M_{j \alpha }^2} \right)}^2}}},
\label{eq:sf}
\end{equation}
which expresses the probability to find a constituent quark $j$ having the momentum fraction $x$ together with a Goldstone boson $\alpha$ inside a parent constituent quark $i$.
In Eq.~\eqref{eq:sf}, $m_i$, $m_j$, and $m_{\alpha}$ are masses of the constituent quarks and the Goldstone boson, respectively, ${\bar m} = (m_i + m_j )/2$ is the average of the constituent quark masses, and $M_{j \alpha }^2$ represents the invariant mass squared of the final state:
\begin{equation}
M_{j \alpha }^2  =  \frac{{m_j^2 + k_T^2}}{y} + \frac{{m_\alpha ^2 + k_T^2}}{{1 - y}}.
\end{equation}
To make the integral in Eq.~\eqref{eq:sf} to converge, in this study we replace $g_A$ with
${g_A} \exp \left[ \left( m_i^2 - M_{j \alpha }^2 \right) / 4 \Lambda ^2 \right]$, where $\Lambda$ is a cutoff.
Using a short hand notation:
\begin{equation}
P \otimes q \equiv \int_x^1 {\frac{{dy}}{y}} P \left( y \right) q\left( {\frac{x}{y}} \right) ,
\end{equation}
the contributions from the corrections described by Fig.~\ref{fig:dressing_corrections}(b) and Fig.~\ref{fig:dressing_corrections}(c) can be expressed as ${P_{j\alpha / i}}\otimes {q}_{i}$ and $V_{k / \alpha}\otimes {P_{\alpha j / i}}\otimes {q}_{i}$, respectively, where  $V_{k / \alpha}$ represents the quark distribution of the Goldstone boson $\alpha$.

In this study, we focus on the valence $u$ and $\bar{s}$ quark distributions of the $K^{+}$ meson.
Collecting all the contributions described by the three diagrams in Fig.~\ref{fig:dressing_corrections}, the dressed valence $u$ quark distribution is written down as
\begin{align}
v^{K (u)}_{\rm dressed} \left( x \right) = &u\left( x \right) - \bar u\left( x \right) \nonumber \\
= &Z_u {u_0}\left( x \right) + \frac{1}{2}{P_{u \pi / u}} \otimes {u_0} + {V_{u / \pi }} \otimes {P_{\pi d/u}} \otimes {u_0} \nonumber \\
&+ {V_{u / K}} \otimes {P_{Ks / u}} \otimes {u_0} + {V_{u / K}} \otimes {P_{K\bar u/\bar s}} \otimes {{\bar s}_0} \nonumber \\
&- {P_{\bar uK / \bar s}} \otimes {{\bar s}_0} + \frac{1}{6}{P_{u\eta / u}} \otimes {u_0}. \label{eq:dressed_uval}
\end{align}
The renormalization constant in Eq.~\eqref{eq:dressed_uval} is obtained as
\begin{equation}
Z_u = 1 - \frac{3}{2}\left\langle {{P_\pi }} \right\rangle  - \left\langle {{P_{K (i = u)}}} \right\rangle - \frac{1}{6} \left\langle {{P_{\eta (i = u)}}} \right\rangle ,
\label{eq:Zu}
\end{equation}
where $\left\langle {{P_{K (i = u)}}} \right\rangle$ and $\left\langle {{P_{\eta (i = u)}}} \right\rangle$ represent the first moments of the splitting functions in which a $u$ quark is the parent constituent quark.
Collecting the contributing terms similarly, the dressed valence $\bar{s}$ quark distribution function can be written down as
\begin{align}
v^{K(\bar{s})}_{\rm dressed} \left( x \right) = &\bar{s}\left( x \right) - s\left( x \right) \nonumber \\
= &{Z_{s}}{{\bar s}_0}\left( x \right) + {V_{\bar s / K}} \otimes {P_{K s / u}} \otimes {u_0} \nonumber \\
&+ {V_{\bar s / K}} \otimes {P_{K\bar u / \bar s}} \otimes {{\bar s}_0} + {V_{\bar s / K}} \otimes {P_{K\bar d / \bar s}} \otimes {{\bar s}_0} \nonumber \\
&+ \frac{2}{3}{P_{\bar s \eta / \bar s}} \otimes \bar s - {P_{s K / u}} \otimes {u_0}, \label{eq:dressed_sbarval}
\end{align}
where the renormalization constant is given by
\begin{equation}
Z_{s} = 1 - 2 \left\langle {{P_{K (i = \bar{s})}}} \right\rangle - \frac{2}{3} \left\langle {{P_{\eta (i = \bar s )}}} \right\rangle . \label{eq:Zsbar}
\end{equation}
%

%%%%%%%%%%%%%%%%%%%%%%%%%%%%%%%%%%%%%%%%
\section{Numerical results}
\label{sec-3}
%%%%%%%%%%%%%%%%%%%%%%%%%%%%%%%%%%%%%%%%
Here we present our numerical results, and discuss the general features of the meson cloud effects that we can see through the analysis.
To numerically evaluate the dressed quark distribution functions of the kaon, we need the bare distributions of the pion and the kaon at the initial scale as inputs.
As to the pion, in this work we utilize the result obtained in the previous study~\cite{Watanabe:2016lto}:
\begin{equation}
v^{\pi}_{\rm bare}(x, Q_0^2) = N_{\pi}x^{\alpha} (1 - x)^{\alpha},
\label{eq:bare_uval_of_pion}
\end{equation}
where $N_{ \pi }$ is the normalization factor, and $\alpha = 1.8$ was determined to reproduce the phenomenologically satisfactory valence quark distribution of the pion at $Q^2 = 16$~GeV$^2$~\cite{Aicher:2010cb}.
On the other hand, as to the kaon we assume the following functional forms for its bare distribution functions:
\begin{align}
&v^{K (u)}_{\rm bare} (x, Q_0^2) = N_{K (u)} x^{\beta} (1 - x)^{\gamma}, \\
&v^{K (\bar{s})}_{\rm bare} (x, Q_0^2) = N_{K(\bar{s})} x^{\gamma} (1 - x)^{\beta},
\label{eq:bare_val}
\end{align}
where $N_{K (u)}$ and $N_{K (\bar{s})}$ represent the normalization factors, and $\beta$ and $\gamma$ are the adjustable parameters.
To determine those parameters, we utilize the experimental data of the valence $u$ quark distribution function ratio of the pion and the kaon, $v^{K(u)} (x, Q^2) / v^{\pi (u)} (x, Q^2)$, which was taken at $Q^2 = 27$~GeV$^2$~\cite{Badier:1980jq}.
Evaluating the dressing corrections and performing the QCD evolution, the calculated ratio is compared with the data at the scale, which enables us to obtain the best fit for the parameters.
For this procedure, the DGLAP evolution code~\cite{Kobayashi:1994hy} is used at NLO accuracy, and the MINUIT package~\cite{James:1975dr} is adopted for the fitting.
The obtained best fit values are $\beta = 1.84 \pm 0.27$ and $\gamma = 2.01 \pm 0.19$.

\begin{figure}[tb]
\centering
\includegraphics[width=0.45\textwidth]{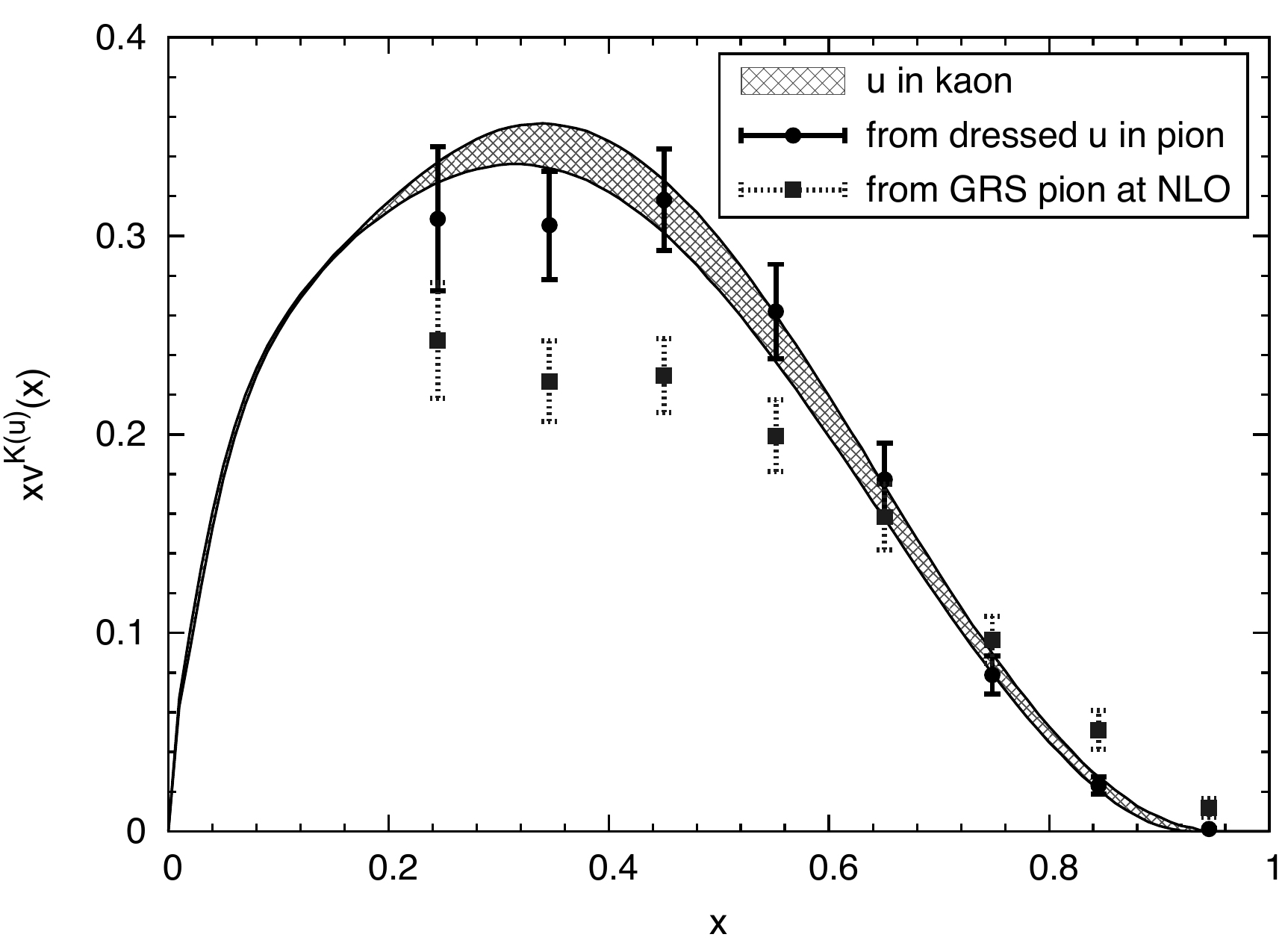}
\caption{
The cross-hatched pattern depicts the 68\% C.L. uncertainty of the resulting valence $u$ quark distribution of the kaon at $Q^2 = 27$~GeV$^2$.
The circles represent the empirical values from the combination of the data of the ratio $v^{K(u)} / v^{\pi(u)}$~\cite{Badier:1980jq} and the pion valence quark distribution function taken from Ref.~\cite{Watanabe:2016lto}.
The squares show the values obtained from the combination of the ratio data and the GRS parametrization~\cite{Gluck:1999xe}.
}
\label{fig:fit_result}
\end{figure}

We display in Fig.~\ref{fig:fit_result}
the resulting valence $u$ quark distribution of the $K^{+}$ meson at $Q^2 = 27$~GeV$^2$, which is depicted by the  cross-hatched band.
The results denoted by the circles and the squares are obtained from the valence quark distributions of the pion, which are taken from Ref.~\cite{Watanabe:2016lto} and the Gl\"uck-Reya-Schienbein (GRS) parametrization~\cite{Gluck:1999xe}, multiplied by the experimental data of the ratio $v^{K (u)} / v^{\pi (u)}$.
It is seen from the figure that the results depicted by the circles are consistent with our result within the errors, which shows that our fitting is successful.
Another important observation is that the results obtained from the GRS parametrization are considerably smaller compared to our calculation at $x < 0.6$.

\begin{figure}[tb]
\centering
\includegraphics[width=0.45\textwidth]{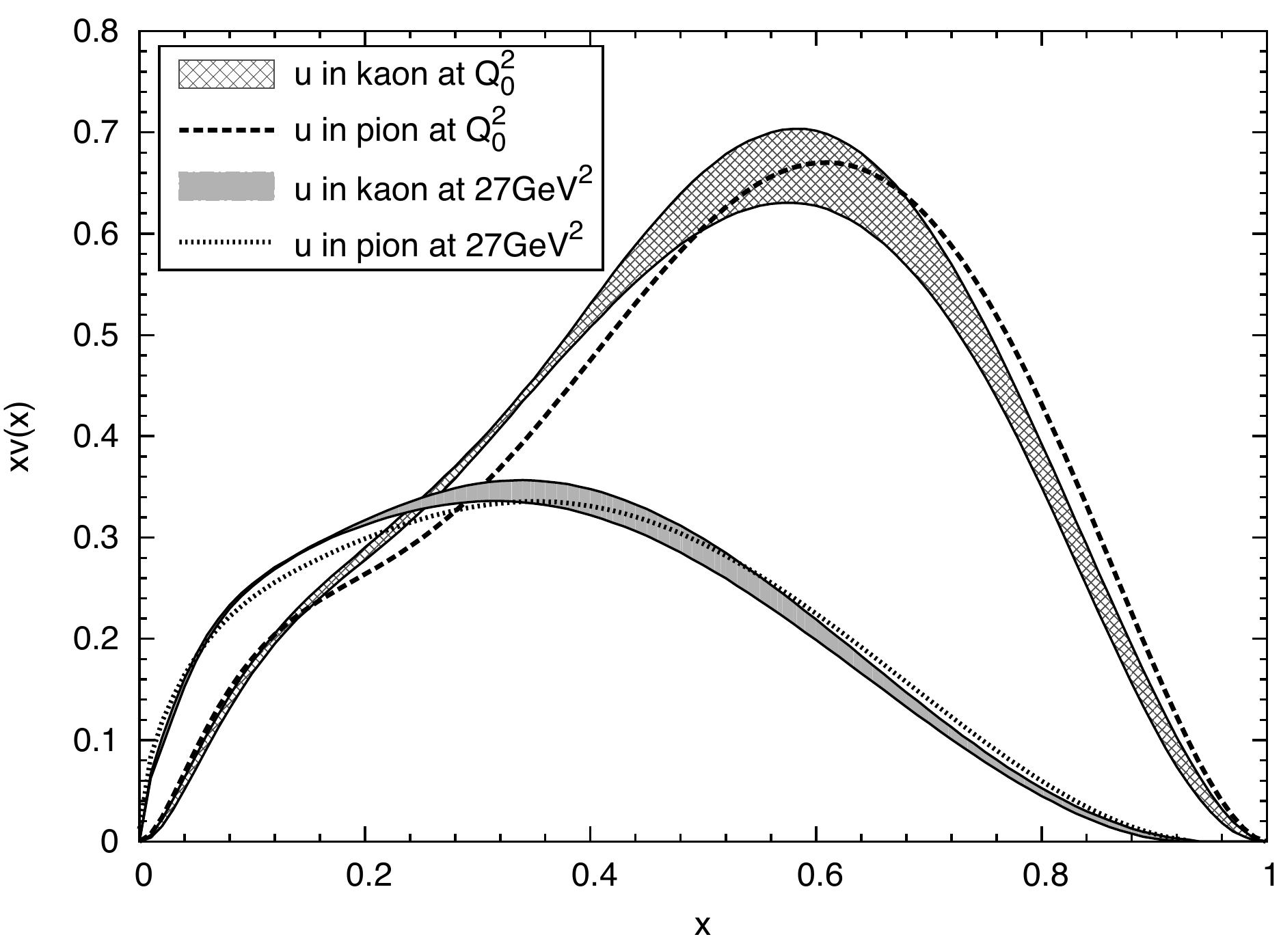}
\caption{
The dressed valence $u$ quark distribution functions of the pion and the kaon as a function of Bjorken $x$.
The cross-hatched and shadowed bands represent the 68\% C.L. $v^{K (u)}_{\rm dressed} (x)$ uncertainties at $Q^2 = Q_0^2$ and $27$~GeV$^2$, respectively.
The dashed and dotted lines depict the distribution functions of the pion taken from Ref.~\cite{Watanabe:2016lto} at $Q^2 = Q_0^2$ and $27$~GeV$^2$, respectively.
}
\label{fig:dressed_uval_with_errors}
\end{figure}

\begin{figure}[tb]
\centering
\includegraphics[width=0.45\textwidth]{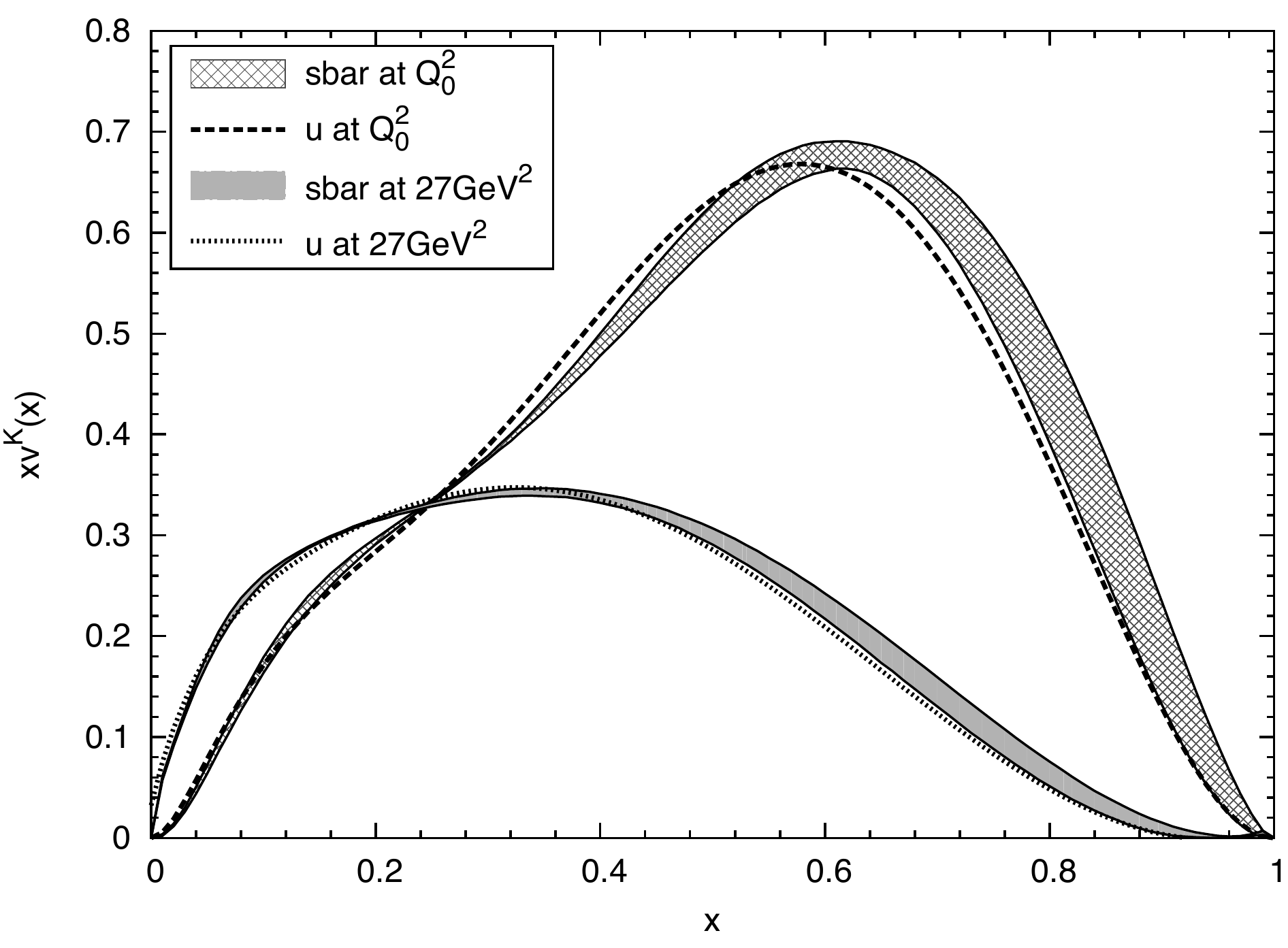}
\caption{
The dressed valence quark distributions of the kaon as a function of Bjorken $x$.
The cross-hatched and shadowed bands represent the 68\% C.L. $v^{K (\bar{s})}_{\rm dressed} (x)$ uncertainties at $Q^2 = Q_0^2$ and $27$~GeV$^2$, respectively.
The dashed and dotted lines depict the valence $u$ quark distribution functions at $Q^2 = Q_0^2$ and $27$~GeV$^2$, respectively.
}
\label{fig:dressed_QDF_in_kaon_with_errors}
\end{figure}

We show in Fig.~\ref{fig:dressed_uval_with_errors}
the comparison between the resulting dressed valence $u$ quark distribution function of the kaon and that of the pion taken from Ref.~\cite{Watanabe:2016lto} at $Q^2 = Q_0^2$ and $27$~GeV$^2$.
From this figure, it is seen that $xv^{\pi}$ is smaller than $xv^{K(u)}$ in the region where $0.2 < x < 0.5$, but larger when $x > 0.7$ at $Q^2 = Q_0^2$.
At $Q^2 = 27$~GeV$^2$, the difference between the two results becomes smaller, and it is hard to find it particularly in the region where $x > 0.3$.
We also present the comparison between the resulting dressed valence $u$ and $\bar{s}$ quark distributions of the kaon at $Q^2 = Q_0^2$ and $27$~GeV$^2$ in Fig.~\ref{fig:dressed_QDF_in_kaon_with_errors}.
In this figure, we show the uncertainties only for the $\bar{s}$ quark results because of the comparison purpose.
One can see that the difference between the two results at the initial scale is tiny and they are almost identical to each other at $Q^2 = 27$~GeV$^2$, although this can also be understood from the fitting results for the parameters, $\beta$ and $\gamma$.

It is useful to consider the moments of the distribution functions for a comparison between our results and those from preceding studies.
The obtained values of the moments are presented in Table~\ref{table:VQDF_moments}.
\begin{table}[b]
\caption{
The first three moments calculated from the resulting dressed valence quark distributions of the kaon at $Q^2 = 27$~GeV$^2$, compared to those from Ref.~\cite{Chen:2016sno}.
}
\begin{center}
\begin{tabular}{| c | c || c | c | c |}
\hline
$q$ & & $\langle x \rangle _q$ & $\langle x^2 \rangle _q$ & $\langle x^3 \rangle _q$  \\
\hline
\hline
$v^{K(u)}$ & This work & 0.23 & 0.091 & 0.045 \\
& \cite{Chen:2016sno} & 0.28 & 0.11 & 0.048  \\
\hline
$v^{K(\bar{s})}$ & This work & 0.24 & 0.096 & 0.049  \\
& \cite{Chen:2016sno} & 0.36 & 0.17 & 0.092 \\
\hline
\end{tabular}
\end{center}
\label{table:VQDF_moments}
\end{table}
For a comparison, we also show the values obtained from a Dyson-Schwinger equation based study~\cite{Chen:2016sno} in the table.
The most important observation here is that the size of the SU(3) flavor symmetry breaking which can be seen from our results is considerably smaller than that seen in the preceding study.

\begin{figure}[tb]
\centering
\includegraphics[width=0.45\textwidth]{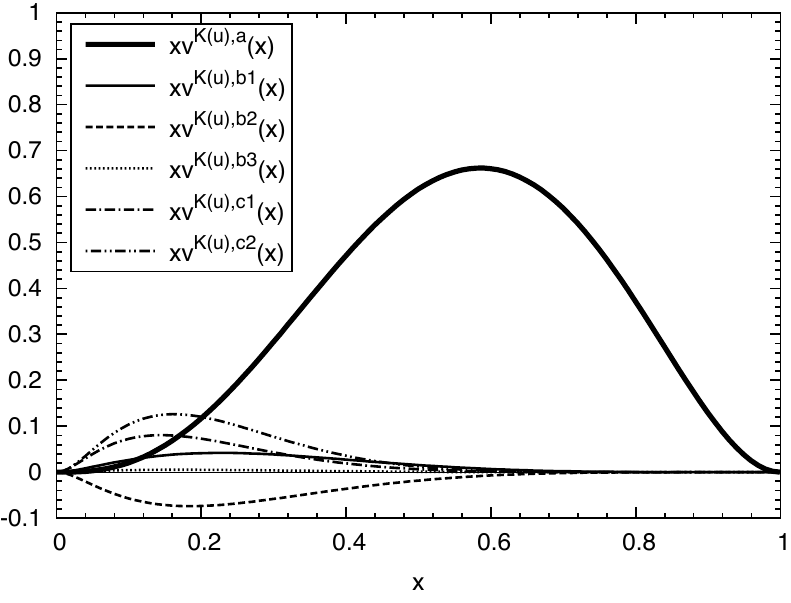}
\caption{
The behaviors of each term in the right-hand side of Eq.~\eqref{eq:deconposition} at $Q^2=Q_0^2$ as functions of Bjorken $x$.
}
\label{fig:term_by_term}
\end{figure}

Next, let us see the details of the meson cloud effects, focusing on the corrections to the valence $u$ quark of the kaon.
To do so, we decompose the dressed distribution, Eq.~\eqref{eq:dressed_uval}, into several terms:
\begin{align}
{v^{K(u)}_{\rm dressed}}\left( x \right) = &{v^{K (u), a}}\left( x \right) + {v^{K (u), b1}}\left( x \right) + {v^{K (u), b2}}\left( x \right) \nonumber \\
&+ {v^{K (u), b3}}\left( x \right) + {v^{K (u), c1}}\left( x \right) + {v^{K (u), c2}}\left( x \right) , \label{eq:deconposition}
\end{align}
where
\begin{align}
&{v^{K (u), a}}\left( x \right) = {Z_u}{u_0}\left( x \right), \ {v^{K (u), b1}}\left( x \right) = \frac{1}{2}{P_{u \pi / u}} \otimes {u_0}, \nonumber \\
&{v^{K (u), b2}}\left( x \right) = - {P_{\bar u K / \bar s}} \otimes {{\bar s}_0}, \ {v^{K (u), b3}}\left( x \right) = \frac{1}{6}{P_{u \eta / u}} \otimes {u_0}, \nonumber \\
&{v^{K (u), c1}}\left( x \right) = {V_{u / \pi }} \otimes {P_{\pi d / u}} \otimes {u_0}, \nonumber \\
&{v^{K (u), c2}}\left( x \right) = {V_{u / K}} \otimes {P_{K s / u}} \otimes {u_0} + {V_{u / K}} \otimes {P_{K \bar u / \bar s}} \otimes {{\bar s}_0} .
\label{eq:term_by_term}
\end{align}
The behaviors of the above terms at the initial scale are displayed in Fig.~\ref{fig:term_by_term}.
The thick solid line corresponding to $xv^{K(u),a}$ shows the contribution from the renormalization constant, which obviously dominates in the region where $x > 0.6$.
Hence, it is understood that the resulting valence $u$ quark distribution function behaves as $(1 - x)^{\gamma}$ when $x\rightarrow 1$.
The contributions depicted by the diagrams in Figs.~\ref{fig:dressing_corrections}(b) and \ref{fig:dressing_corrections}(c) are nonzero only in the region of $x < 0.6$, and only $xv^{K (u), b2}$ gives a negative contribution.
It is observed that $xv^{K (u), c2}$ is larger than $xv^{K (u), c1}$ in the whole $x$ regime, and $xv^{K (u), c1}$ is larger than $xv^{K (u), b1}$ in the region where $x < 0.3$.
The contribution from the $\eta$ meson, $xv^{K (u), b3}$, is extremely small.

\begin{figure}[tb]
\centering
\includegraphics[width=0.45\textwidth]{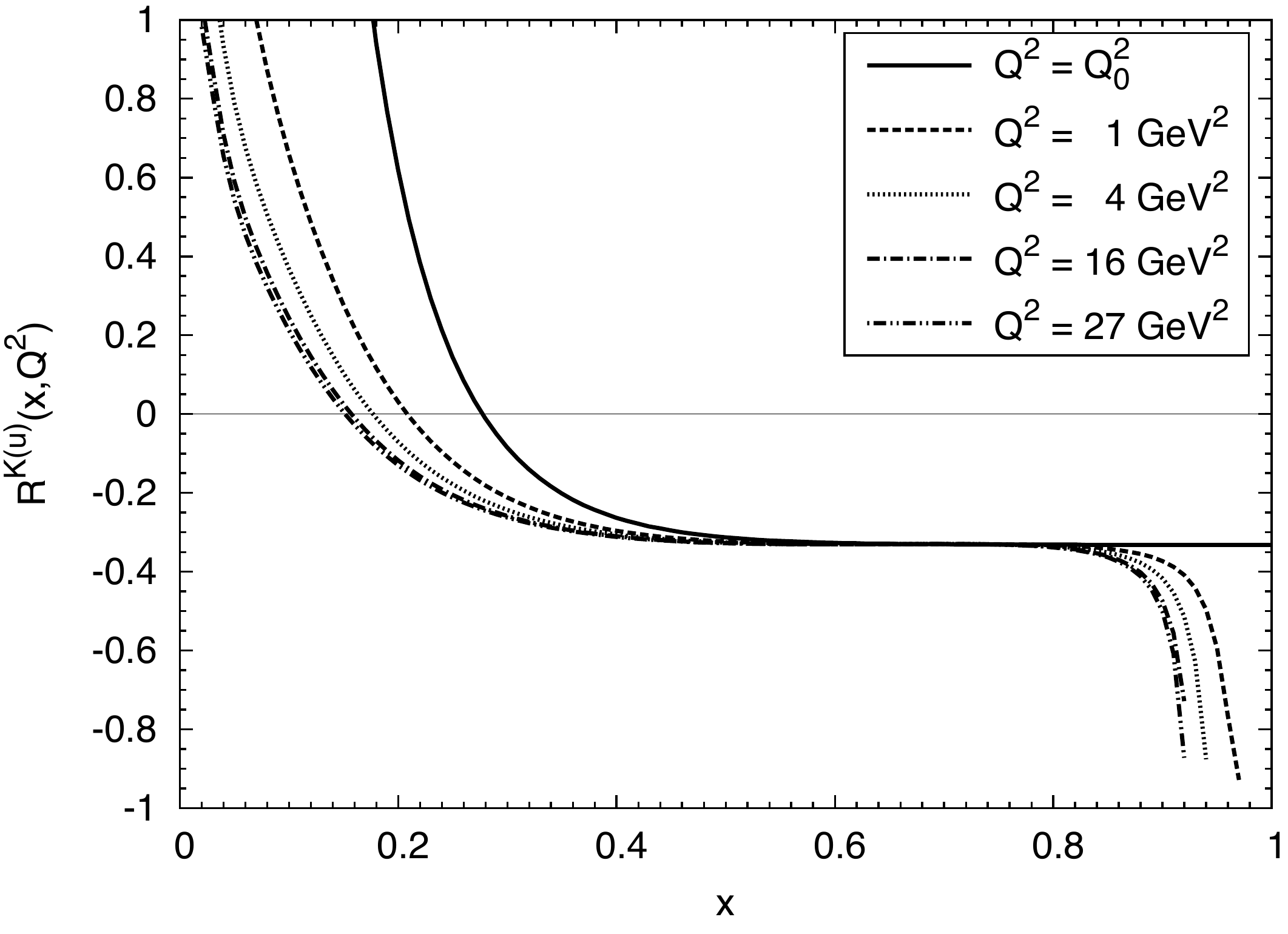}
\caption{
The ratio of the valence quark distribution function, $R^{K (u)}(x, Q^2)$ defined in Eq.~\eqref{eq:R}, as a function of Bjorken $x$ for various $Q^2$.
}
\label{fig:R}
\end{figure}

Finally, we show in Fig.~\ref{fig:R} the $x$ and $Q^2$ dependencies of the valence quark distribution function ratio defined by
\begin{equation}
R^{K (u)} \left( {x, Q^2} \right) \equiv \frac{{v_{{\rm{dressed}}}^{K (u)} \left( {x, Q^2} \right) - v_{{\rm{bare}}}^{K (u)} \left( {x, Q^2} \right)}}{{v_{{\rm{bare}}}^{K (u)} \left( {x, Q^2} \right)}}, \label{eq:R}
\end{equation}
to understand the relative magnitude of the dressing correction.
It is seen from the figure that there is the overall 33\% reduction because of the renormalization constant in the large $x$ region at $Q^2=Q_0^2$, but at higher $Q^2$ the ratio is further suppressed in the region of $x > 0.8$.
This suppression becomes stronger with $Q^2$.
The large $x$ behavior of the valence quark distribution function is usually characterized with some exponent $\delta$ by $v (x)\sim (1 - x)^{\delta}$ when $x \sim 1$.
Therefore, it is understood that the meson cloud effect makes the value of $\delta$ to be larger with $Q^2$.
On the other hand, in the small $x$ region the effect enhances the bare distribution when $x < 0.4$, and the enhancement magnitude decreases as $Q^2$ increases.

%

%%%%%%%%%%%%%%%%%%%%%%%%%%%%%%%%%%%%%%%%
\section{Summary}
\label{sec-4}
%%%%%%%%%%%%%%%%%%%%%%%%%%%%%%%%%%%%%%%%
In this work, we have investigated the valence $u$ and $\bar{s}$ quark distribution functions of the $K^{+}$ meson, taking into account the meson cloud effects in the chiral constituent quark model.
Assuming the functional forms of the bare distribution functions and evaluating the dressing corrections, the realistic dressed distributions are obtained.
By utilizing the experimental data of the valence $u$ quark distribution function ratio $v^{K (u)} / v^{\pi (u)}$, the two adjustable parameters included in the assumed bare distributions can be determined, although the data have large uncertainties.
We have presented the resulting dressed distributions and also discussed the general features of the meson cloud effects.

We have found that the observed size of the SU(3) flavor symmetry breaking is considerably smaller compared to the results obtained from the preceding studies based on other approaches.
Since the presently available data are not enough to pin down this, more precise data are certainly needed.
The kaon-induced Drell-Yan experiment at some facilities such as COMPASS at CERN and J-PARC, utilizing the high intensity kaon beam, is expected to be carried out in the future.
Our predictions presented here can be tested there.

\nocite{*}
\bibliographystyle{elsarticle-num}
\bibliography{kaon_QDF_in_CCQM}

\end{document}